\documentclass{ws-mplb}
\usepackage[numbers,super]{natbib}  
\makeatletter
\DeclareRobustCommand\onlinecite{\@onlinecite}
\def\@onlinecite#1{\begingroup\let\@cite\NAT@citenum\citealp{#1}\endgroup}
\makeatother

\begin{document}

\title{\Large\bf  Superconductivity near Pomeranchuk instabilities in
the spin channel}
\author{Daniel G.\ Barci}
\address{Departamento de F{\'\i}sica Te\'orica,
Universidade do Estado do Rio de Janeiro, Rua S\~ao Francisco Xavier 524, 20550-013,  Rio de Janeiro, RJ, Brazil.}
\author{Paulo S. A. Bonfim}
\address{Universidade UnigranRio, Rua Prof. Jos\'e de Souza Herdy, 1.160, 25071-202, Duque de Caxias,RJ, Brazil}

\maketitle

\begin{abstract}
We study the competition between a Pomeranchuk instability in the spin channel with angular 
momentum $\ell=1$ and an attractive interaction, favoring Cooper-pair formation.
We found that the superconducting gap strongly suppresses the phase space for the Pomeranchuk instability.     
We computed  a  mean-field phase diagram displaying a first order transition between two superconductor phases with different
symmetries: p-wave (with spontaneously generated spin-orbit interaction) and
s-wave for greater values of the coupling constant. Moreover, we have looked for
a possible modulated superconducting phase. We have found that this
phase appears only as a meta-stable state in the strong coupling regime.  
\end{abstract}

\keywords{Fermi liquid instabilities; Pomeranchuk; Superconductivity}


  
\section{Introduction}
A Fermi liquid is, except in one dimension, a very stable state of matter\cite{nozieres-1999}. 
At least two types of instabilities, related with attractive interactions, are known: Pomeranchuk\cite{pomeranchuk-1958} and superconducting instabilities. 
Pomeranchuk instabilities occur in the presence of  two-body interactions containing a strong attractive component in the forward scattering channel 
with definite angular momentum. In the context of Landau theory, the instability
sets in when one or more  dimensionless Landau parameters $F_{\ell}^{s,a}$,
with angular momentum $\ell$ in the charge ($s$) or spin ($a$) channel, acquire
large negative values. 

Pomeranchuk instabilities in a charge sector spontaneously break rotational
symmetry\cite{Hiroyuki}. In particular, an instability in the 
$F_2^s$ channel produces  an ellipsoidal deformation of  the Fermi surface\cite{oanesyan-2001,barci-2003}. From a dynamical point of view,
the resulting anisotropic ground state is a non-Fermi liquid, due to over-damped Goldstone modes that wipe out quasi-particle 
excitations\cite{lawler-2006,Metlitski-2010,barci-2008}.  

Instabilities in the spin channel are also very
interesting.  For instance,  the ferromagnetic Stoner instability\cite{zhang-2005} occurs when $F_0^a$ acquires 
large negative values, producing a divergence in the magnetic susceptibility. 
This phase transition preserves rotational symmetry, however it breaks time-reversal symmetry.
On the other hand, it is possible to have spin instabilities that preserve time-reversal invariance\cite{Hirsch,cabra-2012}.
Higher order angular momentum interactions produce anisotropic as well as isotropic phases. 
Several examples were studied in detail\cite{wu-2004,wu-2007}. The $\ell=1$
channel have special interest.  When $F^a_1<0$, an ordered isotropic and
time reversal invariant phase is possible. 
This phase, called $\beta$-phase in 
ref. [\onlinecite{wu-2004}],  dynamically generates a spin-orbit coupling. This
is a very interesting possibility, since it allows the generation of spin-orbit
couplings from many-body correlations,  differently from the usual one-particle
relativistic effect.

On the other hand, superconductivity in the presence of spin-orbit interactions is nowadays of great interest\cite{Sigrist-2004}.
Superconductivity is developed in the presence of a small attractive interaction in the particle-particle (BCS) channel. 
The superconducting (SC) state is generally characterized by a complex order parameter which 
breaks gauge symmetry, $\Delta_{\sigma,\sigma'}(\vec r,\vec r')=\langle\psi^{\dagger}_\sigma(\vec r)\psi^{\dagger}_{\sigma'}(\vec r')\rangle$, 
where the operator $\psi^{\dagger}_\sigma(\vec r)$ creates an electron with spin $\sigma$ at the position $\vec r$.
The usual classification of $\Delta_{\sigma,\sigma'}(\vec r,\vec r')$ as s-wave,
d-wave, p-wave, etc, resides in the irreducible representations of the lattice
point group. Also, the absence of spin-orbit interactions allows the additional
differentiation between singlet and triplet order parameters.  However, the SC
state could also break lattice translation and/or rotational symmetry. In that
case, this classification is no more possible. One particular example is an
oscillating order parameter like $\Delta_{\sigma,\sigma'}(\vec
r,0)=\Delta_{\sigma,\sigma'}\cos(\vec q\cdot \vec r)$, where $\vec q$ is an
ordering wave-vector.
In a  modulated superconducting state, proposed by Fulde, Ferrell, Larkin and  Ovchinnikov\cite{fulde-1964,larkin-1964} (FFLO state), the spatial modulation of the 
Cooper pairing is due to a mismatch of Fermi surfaces, produced  by an external magnetic field (by Zeeman effect).
FFLO states  have also been proposed to occur in other scenarios like in imbalanced  cold atoms with different species\cite{radzihovsky-2008} and in heavy fermions systems when  different orbitals hybridize under external pressure\cite{padilha-2009}. 
Also,  for generally non-local BCS potentials, a modulated SC order parameter can coexist with charge density waves\cite{loder-2009}. 
Recently, a striped order parameter, called ``Pair Density Wave, (PDW)'',  was proposed\cite{berg-2007,berg-2009,berg-2009b,barci-2011} 
to explain anomalous transport properties observed in cuprates superconductors.

Although Pomeranchuk and BCS instabilities are 
generated by attractive interactions, they are competing ones.  
A superconducting gap suppresses Fermi surface deformations. A detailed example of this effect can be 
observed by considering      
the competition between d-wave Pomeranchuk instabilities and a 
d-wave superconducting order parameter\cite{metzner-2000,metzner-2007}.

In the spin channel, due to time-reversal symmetry breaking,  a stronger competition is expected. 
However, it is possible to have instabilities with higher angular momentum that preserve time-reversal invariance. 
In this paper, we analyze an example of this class of systems. 
In particular, the $\beta$-phase\cite{wu-2004} 
opens the possibility of the formation of Cooper pairs with zero helicity  and finite  momentum, possibly producing a modulated superconducting state. 

\begin{figure}
\begin{center}
\includegraphics[scale=0.4]{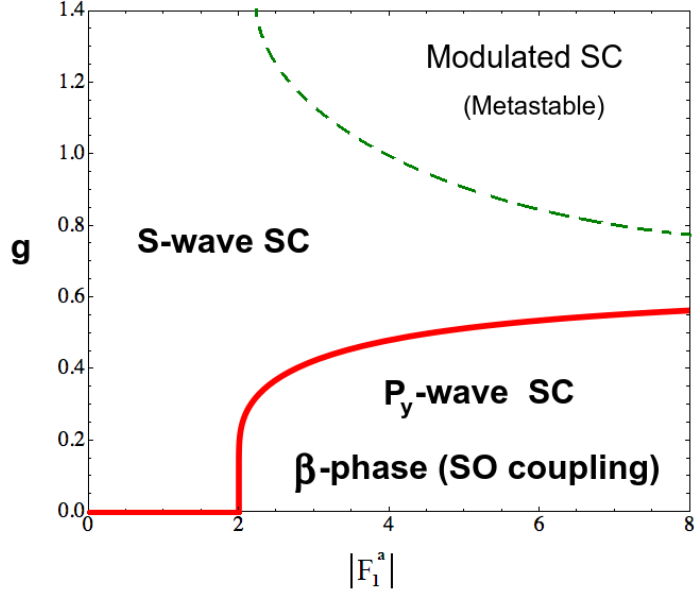}
\caption{Phase diagram in terms of   the BCS coupling constant $g$ and  the dimensionless antisymmetric Landau parameter with angular momentum
$\ell=1$,  $F_1^a=f_1^a N(0)$.    Above the solid line,   the system is mainly in an s-wave  superconducting state ($\bar n=0$, $\Delta_0\neq 0$, $\Delta=0$).  
Below this line, for $F_1^a<-2$ and for small $g$, the system continuously develops an spontaneous spin-orbit interaction, given by the order parameter $\bar n$. 
 The s-wave order parameter goes to zero discontinuously at this line, where  a superconducting order  with $p_y$-wave symmetry, coexisting with a spin-orbit coupling ($\beta$-phase) appears. In this phase $\bar n\neq 0$, $\Delta_0= 0$, $\Delta\neq 0$.  For stronger couplings, the dashed line
indicates the onset of a metastable modulated SC phase ($\Delta_q\neq 0$).}
\label{pd}
\end{center}
\end{figure}

The main point of this paper is to report on the competing character of Cooper
pairing and a Pomeranchuk instability in the $\ell=1$ spin channel. 
Our model is based on a  2D Fermi liquid with interactions in the $F^a_1$
channel and a general attracting interaction favoring Cooper pairing.
The main results are summarized in the phase diagram of figure  (\ref{pd}),
obtained in a self-consistent mean-field approximation described below. 
For $F^a_1> -2$, the system is in the usual s-wave SC state. 
When $F^a_1<-2$ the Pomeranchuk instability sets in and 
we observe two types of superconducting phases. At low coupling $g$, there is a
splitting of the Fermi surface into two branches, labeled by  helicity
eigen-values and, the mean field solution is a p-wave superconducting state
coming from intra-band pairing. Notice that the pairing is in the helicity basis, since the spin in no longer 
a good quantum number. The use of the term p-wave refers to the symmetry of the superconducting gap obtained.  
For greater coupling, Pomeranchuk
instability is suppressed continuously and  the  s-wave superconducting state reappears discontinuously at the bold line. 
Therefore, at mean field level (and zero temperature) there is a  first order phase transition between two different SC states. 
Moreover, for a stronger BCS coupling,  a
modulated superconducting phase could appear with a finite  wave-vector $\vec
q$, given roughly by $|\vec q|\sim k_F\sqrt{1-2/|F_1^a|}$,  produced by 
inter-band BCS couplings. However, this phase is metastable at mean field
level. 

In the rest of the paper we present our model and sketch the mean-field calculation, leading to the phase diagram of fig. (\ref{pd}). 

\section{Model Hamiltonian}
The $\beta$-phase order parameter is given by
\begin{equation}
V^{a\mu}(x)=-i\psi^\dagger(x) \sigma^\mu \hat\nabla_a\psi(x)\;, 
\end{equation}
where $\psi$ is a two-component spinor, $\sigma^\mu$ are the Pauli matrices, and $a=x,y$.
$\hat \nabla e^{ikx}= i \frac{\vec k}{|k|} e^{ikx}$. 

Following ref. [\onlinecite{wu-2004}],
we write a Hamiltonian with forward two-body interactions in this channel
\begin{eqnarray}
H&=&\int d^2x\; \psi^\dagger(x)\left( \epsilon(-i\nabla)-\mu-h\vec\sigma\cdot\vec\nabla  \right)\psi(x) +  \nonumber \\
&+& \frac{1}{2}\int d^2xd^2x' f_1^a(x-x')V^{a\mu}(x)V^{a\mu}(x')\;.
\end{eqnarray} 
The Fourier transform of $f_1^a(x)$ is given by   
$f_1^a(k)=f_1^a/(1+\kappa |f_1^a| k^2)$, defining, in this way, an effective interaction range $r=\sqrt{\kappa |f_1^a|}$.
We have considered,  in the quadratic part of the Hamiltonian, an explicit spin-orbit (SO) coupling with strength $h$. This term will serve to compute the SO susceptibility and we may eventually take 
$h\to 0$. It will also be useful to control the stability of the Pomeranchuk $\beta$-phase. 
For a particle-hole symmetric system we consider the following expansion of  the dispersion relation around a circular Fermi surface, 
\begin{equation}
\epsilon(k)-\mu=\vec v_F\cdot [\vec k-\vec k_F]+ 
\frac{b}{(v_Fk_F)^2} \left(\vec v_F\cdot [\vec k-\vec k_F]\right)^3\;,
\label{expansion}
\end{equation}
where the   dimensionless parameter $b$ measures the effective curvature of the band near the Fermi surface 
and we have ignored terms proportional 
to $(k-k_F)^5$.  The band curvature in the dispersion relation is essential to stabilize any Pomeranchuk instability\cite{barci-2003}, 
for this reason, the usual linear approximation is not suitable to study this type of phase transitions. 

The $\beta$-phase is defined by the mean-field Hamiltonian 
\begin{equation}
H_{\beta}=\int \frac{d^2k}{(2\pi)^2}\; \psi^\dagger(k)\left( \epsilon(k)-\mu-(\bar n+h) \vec\sigma\cdot \hat k \right)\psi(k), 
\end{equation} 
where $\bar n$ is determined self-consistently by
\begin{equation}
\bar n=-\frac{1}{2}f_1^a(0) \int \frac{d^2k}{(2\pi)^2} \langle \psi^\dagger(k)(\vec\sigma\cdot \hat k )\psi(k)\rangle\;.
\label{self-consisten-n}
\end{equation}
This mean-field theory is valid when $k_F\sqrt{\kappa|f_1^a|}\gg 1$, {\em i.\
e.\ },  when the range of the interaction is much larger than the inter-particle
distance. 

\begin{figure}
\begin{center}
\includegraphics[scale=0.4]{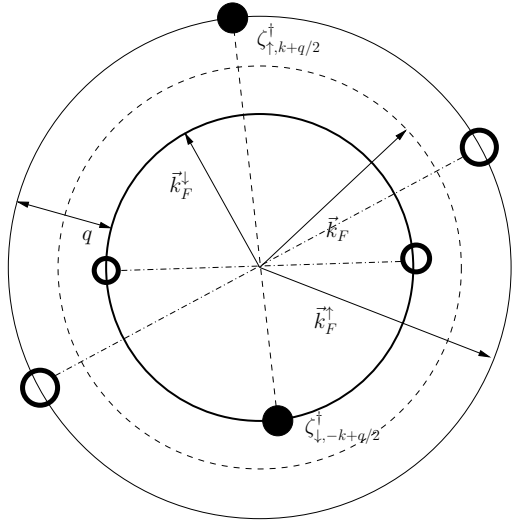}
\caption{Fermi surface splitting in the helicity basis and pairing in the $\beta$-phase. The full dots indicate inter-band  Cooper pairing
with momentum $q$, while the empty dots represent intra-band p-wave Cooper pairing. }
\label{fig.FS}
\end{center}
\end{figure}

It is not difficult to solve eq. (\ref{self-consisten-n}), by considering a Fermi liquid ground state\cite{wu-2004,wu-2007}. 
When $\lim_{h\to 0}\bar n(h)\neq 0$,  the spectrum of $H_{\beta}$ spontaneously   splits into two opposite
chiralities with dispersions $\epsilon^{\uparrow}=\epsilon(k)-(\mu+\bar n)$
and $\epsilon^{\downarrow}=\epsilon(k)-(\mu-\bar n)$, as shown in figure
(\ref{fig.FS}).
The Fermi momentum of each branch is given by $k_F^{\uparrow\downarrow}=k_f\pm
q/2$ and the relation between $q$ and $\bar n$ is computed using
eq. (\ref{expansion}), 
\begin{equation}
\frac{v_F q}{2}=\bar n -\frac{b}{(v_Fk_F)^2}\;  \bar n^3+ O(\bar n^5)\; , 
\label{eq.q}
\end{equation}
which is valid close to the Pomeranchuk instability. To obtain this relation we have fixed the chemical potential. A different result is obtained 
by considering, instead, a fixed density\cite{wu-2007}.

The ordered phase is characterized by a spontaneously  generated spin-orbit coupling with  the global rotation invariance unbroken.
The spin and orbital angular momentum are not conserved independently, however the total angular momentum $\vec J=\vec L+\vec S$ is conserved. 
In the helicity basis $\zeta=(\zeta_{\uparrow},\zeta_{\downarrow})$, where the
operator $\hat k\cdot \vec \sigma$ is diagonalized with eigenvalues $\pm 1$, the
mean-field Hamiltonian takes the simpler form 
\begin{equation}
H_{\beta}=\int \frac{d^2k}{(2\pi)^2}\; \zeta^\dagger(k)\left( \epsilon(k)-\mu-(\bar n+h) \sigma^z \right)\zeta(k)\; , 
\label{MF}
\end{equation} 
with the self-consistent equation
\begin{equation}
\bar n=-\frac{1}{2}f_1^a(0) \int \frac{d^2k}{(2\pi)^2} \langle \zeta^\dagger \sigma^z\zeta\rangle\; . 
\label{n0}
\end{equation}
Therefore, if $\bar n\neq 0$ (with $h=0$), there is a dynamically generated spin-orbit
coupling. The mean field Hamiltonian studied in this example is similar to the
2D Rashba spin-orbit coupling,  characterized by a ground state in which the
spin configuration has a vortex structure in momentum space\cite{wu-2007} with
winding number $w=1$. In the Pomeranchuk transition it is also possible to
obtain a Dresselhaus spin-orbit coupling in which the vortex structure has
winding number $w=-1$. However, both couplings produce the same  splitting
of the Fermi surface as described in fig. (\ref{fig.FS}).  Therefore, the study
of the superconducting instability is completely analog and the phase diagram
is the same of figure (\ref{pd}) in both cases. 
  
A BCS type interaction in the $\psi,\psi^\dagger$ basis induces different types
of interactions in the $\beta$-phase, where the usual Fermi liquid properties
are lost. In the helicity basis $\zeta,\zeta^\dagger$, intra-band p-wave pairing,
 as well as inter-band finite momentum pairing are induced, as shown in figure (\ref{fig.FS}). 
These interactions are competing with each other and with  interactions in the
forward scattering (Pomeranchuk) channel.  

The inter-band pairing can be modeled by considering the
following attractive interaction favoring Cooper-pair formation with linear
momentum $\vec q$,
\begin{eqnarray}
H_{\rm Inter}&=&\int \frac{d^2k}{(2\pi)^2} \Delta_q
\;\zeta^\dagger_{\uparrow,k+q/2} \zeta^{\dagger}_{\downarrow,-k+q/2}+ {\rm
h.c.}\; , 
\label{HSC}
\end{eqnarray}
Note that $\vec q$ (eq. (\ref{eq.q})) is given by the Pomeranchuk instability and should be determined self-consistently.  The ``gap equation'' that complements eq. (\ref{HSC}) is 
\begin{equation}
\Delta_q =-g\int \frac{d^2k}{(2\pi)^2} \langle \zeta_{\downarrow,-k+q/2} \zeta_{\uparrow,k+q/2} \rangle \; .
\label{gapequation0}
\end{equation}

On the other hand, intra-band pairing can be studied by considering 

\begin{eqnarray}
H_{\rm Intra}&=&\int \frac{d^2k}{(2\pi)^2} \left\{\Delta_\uparrow(k)
\;\zeta^\dagger_{\uparrow,k}\zeta^\dagger_{\uparrow,-k} +\Delta_\downarrow(k)
\;\zeta^\dagger_{\downarrow,k}\zeta^\dagger_{\downarrow,-k}         \right\}  
+ {\rm
h.c.}\; , 
\label{Intra}
\end{eqnarray}
where the
simplest
representation
for the gap
function is
$\Delta_{\uparrow\downarrow}(k)=\Delta_{\uparrow\downarrow}\sin(\theta)$ with
$(k_x,k_y)=k(\cos\theta, \sin\theta)$, in such a way that
$\Delta_{\uparrow\downarrow}(k)=-\Delta_{\uparrow\downarrow}(-k)$. The self
consistent constraint that complements this mean field Hamiltonian is 

\begin{equation}
\Delta_{\uparrow\downarrow} =-g\int \frac{d^2k}{(2\pi)^2} \sin\theta\; \langle
\zeta_{\uparrow\downarrow,k} \zeta_{\uparrow\downarrow,-k} \rangle \; .
\label{gapequationpw}
\end{equation}
Thus, $\Delta_{\uparrow\downarrow} $ is the order parameter of a $p_y$-wave SC phase. 
The $p_x$ component, proportional to $\cos\theta$, exactly cancels due to the anti-commuting character of the fermionic operators in Eq. (\ref{Intra}). 
This is due to the simple  symmetry of the form factors used to define the BCS couplings.   

\section{Self-consistent solution}
To find the phase diagram of figure (\ref{pd}) we need to solve
the self-consistent equations (\ref{n0}), (\ref{gapequation0}) and 
(\ref{gapequationpw}) where 
the expectation values should be computed with the mean-field Hamiltonian 
$H=H_{\beta}+H_{\rm Intra}+H_{\rm Inter}$ given by equations (\ref{MF}), 
(\ref{HSC}) and (\ref{Intra}).  

We used the Green's function method to compute expectation values, 
\begin{equation}
\langle B A\rangle=\int d\omega f(\omega) Im\left(\ll B;A\gg\right)
\label{expectation}
\end{equation}
where $A$ and $B$ are two fermionic operators and $f(\omega)$ is the Fermi function. 
The notation $\ll A;B\gg$ indicates the Green's function associated with these operators and can be computed 
using the equation of motion for Green's functions which, in frequency
representation, reads\cite{Zubarev}, 
\begin{equation}
\omega \ll A;B\gg=\ll [A,H];B\gg+\langle\left\{A,B\right\}\rangle
\label{eqMotion}
\end{equation}
Since $H$ is quadratic, equation (\ref{eqMotion}) has closed solutions. We can compute for instance, 
\begin{equation}
\ll  \zeta_{\uparrow,k}; \zeta^\dagger_{\uparrow,k}\gg=
\frac{\left(\omega+\epsilon_k^\uparrow\right)\left(\omega+\epsilon_{k-q}^\downarrow\right)}
{\left(\omega+\epsilon_{k-q}^\downarrow\right)  \left[\omega^2-(\epsilon_k^\uparrow)^2-\Delta^2(\theta) \right]
-\Delta_q^2 \left(\omega+\epsilon_k^\uparrow\right)}\; , 
\label{GF}
\end{equation}
where, for simplicity, we assumed $\Delta_\uparrow=\Delta_\downarrow=\Delta$.
Of course, this procedure is completely equivalent to the diagonalization of the Hamiltonian by means of a Bogoliuvov transformation. 

Single particle excitations are given by the poles of the Green's function. In our case this is a cubic polynomial and can be easily computed.  
For weak couplings, {\em i. e. }, for infinitesimal BCS coupling, we cannot have
modulated superconductivity\cite{loder-2009}. Therefore, in the region where
the 
Pomeranchuk instability sets in, and for small values of the coupling constant, the only possible solution of equation (\ref{gapequation0}) is $\Delta_q=0$. 
Therefore, in this regime, both helicity branches are decoupled, and single particle excitations have the dispersion relation 
\begin{equation}
\omega^2=\left(\epsilon^{\uparrow\downarrow}_ k\right)^2+\Delta^2(\theta).
\end{equation}   

Taking the imaginary part of the Green's function and using equation (\ref{expectation}) to compute expectation values, we can rewrite equation 
(\ref{n0}) as 
\begin{eqnarray}
\bar n&=& -\frac{1}{2}f_1^a(0) \int \frac{d^2k}{(2\pi)^2} \left(\langle \zeta_\uparrow^\dagger \zeta_\uparrow\rangle-
\langle \zeta_\downarrow^\dagger \zeta_\downarrow\rangle 
\right) \nonumber \\
&= & -\frac{1}{2}f_1^a(0) 
\int \frac{d^2k}{(2\pi)^2} \left(  
\frac{\epsilon_k^\uparrow}{\sqrt{(\epsilon_k^\uparrow)^2+\Delta^2(\theta)}}- 
\frac{\epsilon_k^\downarrow}{\sqrt{(\epsilon_k^\downarrow)^2+\Delta^2(\theta)}}
\right) 
\label{n1}
\end{eqnarray}
Performing the integrals, and considering negative values of $f_1^a(0)$ we find, 
\begin{equation}
\bar n= \frac{1}{2}|F_1^a| \left(1+8\left(\frac{\pi\Delta}{2\Lambda}\right)^2\ln\left(\frac{\pi\Delta}{2\Lambda}\right)\right)\bar n\left[1-\frac{b}{(v_Fk_F)^2}\bar n^2 \right]
\label{n2}
\end{equation}
with $F_1^a=f_1^aN(0)$, where $N(0)$ is the density of states at the Fermi
surface, and $\Lambda$ is an ultraviolet momentum cut-off. 
 In the absence of pairing $\Delta=0$, this equation has non trivial solutions $\bar n\neq 0$ for $F_1^a<-2$, signaling the usual Pomeranchuk instability. 
 The presence of pairing restricts this region to 
 \begin{equation}
 |F_1^a|> \frac{2}{1+8\left(\frac{\pi\Delta}{2\Lambda}\right)^2\ln\left(\frac{\pi\Delta}{2\Lambda}\right)}
 \end{equation}

On the other hand, within this region, the two helicity branches are decoupled
and the gap equation for $p_y$-wave pairing, equation (\ref{gapequationpw}) reads, 
\begin{equation}
\Delta =-g\int \frac{d^2k}{(2\pi)^2}  \frac{\Delta \sin^2\theta}{\sqrt{\epsilon_k^2+\Delta^2 \sin^2\theta}}
\label{gapequationpw2}
\end{equation}
which, due to the usual logarithmic divergence, has a non-trivial solution  $\Delta\sim \Lambda e^{-1/g}$.
Therefore, at mean-field level,  the system develops a p-wave superconductivity coexisting with a spin-orbit interaction in a small region of the phase 
diagram under the bold line in figure (\ref{pd}). 
As can be seen from equation (\ref{n2}), the line 
\begin{equation}
 |F_1^a|=\frac{2}{1+2\pi^2e^{-2/g}\ln\left[(\pi/2) e^{-1/g}\right]}
 \label{PL}
 \end{equation}
 represents a continuous Pomeranchuk transition between 
$\bar n\neq 0$ and $\bar n=0$. Above this transition, there is no more splitting
of the Fermi surface, since $q=0$, and the character of the ground  state changes
qualitatively.  

Above the Pomeranchuk line, Eq. (\ref{PL}),  the  one-particle excitations are given now by, 
\begin{equation}
\omega^2=\epsilon_ k^2+\Delta_0^2+\Delta^2\sin^2\theta
\end{equation}
as can be seen from the structure of the Green's function, Eq. (\ref{GF}), by taking $q\to 0$.
In this regime, the gap equations  (\ref{gapequation0}) and (\ref{gapequationpw}) are coupled, 
\begin{eqnarray}
\Delta &=&-g\int \frac{d^2k}{(2\pi)^2}  \frac{\Delta \sin^2\theta}{\sqrt{\epsilon_k^2+\Delta^2_0+\Delta^2 \sin^2\theta}}   
\label{gapequationpw+}  \\
\Delta_0 &=&-g\int \frac{d^2k}{(2\pi)^2}  \frac{\Delta_0}{\sqrt{\epsilon_k^2+\Delta^2_0+\Delta^2 \sin^2\theta}}   
\label{gapequationpw-}.
\end{eqnarray}
The only solution of these equations above the curve $g(F_1^a)$ (Eq. (\ref{PL})) is $\Delta=0$, $\Delta_0\sim e^{-1/g}$.  For this reason, the bold curve in the phase diagram
represents  a first order transition between two superconducting states: the
``normal'' s-wave superconductor and a $p_y$-wave superconductor state characterized
by a spin-orbit coupling responsible for the helicity splitting of the Fermi
surface.  
Note that, despite that  pairing is taking place in the helicity basis,  we use the  terms s-wave and p-wave to denote  different symmetries of order parameters.
We stress that, the $p_y$-wave order parameter in the $\beta$-phase, when written in terms of the original Cartesian basis, is a mixture with other partial wave channels, including s-wave.

The only interesting possibility we have  left aside is the  presence  of  modulated superconductivity due to the spontaneous Fermi surface splitting,  produced by
a  Pomeranchuk instability.  In the next subsection we analyze this possibility. 

\subsection{Looking for modulated superconductivity}
We would like to look for a solution of the self consistent equations of the form $\bar n\neq 0$, $\Delta_q\neq 0$ and $\Delta=0$. 
In this case, the single particle excitations are,
\begin{equation}
\omega_{\pm}=
\left(\frac{\epsilon^\uparrow_{k+q/2}-\epsilon^\downarrow_{k-q/2}}{2} \right)
\pm \xi  \; ,
\label{omega}
\end{equation}
with
\begin{equation}
\xi=  \sqrt{\left(\frac{\epsilon^\uparrow_{k+q/2}+\epsilon^\downarrow_{k-q/2}}{2}\right)^2+|\Delta_q|^2}\; .
\label{xi}
\end{equation}

The gap equations (\ref{n0}) and (\ref{gapequation0}) are written now as
 (at zero temperature)
\begin{equation}
\bar n=\frac{|f_1^a|}{4} \int \frac{d^2k}{(2\pi)^2} \left\{\Theta(-\omega_+)-\Theta(\omega_-)\right\}\; , 
\label{n}
\end{equation}
\begin{equation}
\Delta_q=-g \int \frac{d^2k}{(2\pi)^2} \frac{\Delta_q}{2\xi}\left\{\Theta(-\omega_+)-\Theta(-\omega_-)\right\} \; , 
\label{Delta}
\end{equation}
where  $\Theta$ is the usual Heaviside function.
It is simple to realize that the s-wave BCS order parameter $\bar n=0, \Delta_q=\Delta_0$ is always a solution of eqs. 
(\ref{n}) and (\ref{Delta}).     

Now, we explore the interesting possibility of a modulated solution $\bar n\sim q\neq 0, \Delta_q\neq 0$.
The $\vec k$-integrals in equations (\ref{n}) and (\ref{Delta}) are strongly constrained by the Heaviside functions.
The main contribution to eq. (\ref{n}) comes from the region $\omega_+<0$.
Written in polar coordinates $(k,\theta)$, with $\cos\theta=(\vec k\cdot
\vec q)/kq$, and first integrating  over $k$, we get
\begin{eqnarray}
\lefteqn{
\bar n= |F^a_1|\bar n\int_{\theta^-}^{\theta^+} \frac{d\theta}{2\pi}\times } \nonumber \\
&\times& \sqrt{\left(1-\frac{b}{(v_Fk_F)^2}\bar n^2 \cos\theta\right)^2(1+\cos\theta)^2-x^2}
\end{eqnarray} 
where we have defined the dimensionless quantity $x=\Delta_q/\bar n$.
$\theta^{\pm}(x,\bar n)$ are the integration limits which keep the argument of
the square root positive.  $|F^a_1|= N(0) f^a_1$, where $N(0)$ is the density of states at the original Fermi surface (with $\bar n=0$). 
This approximation is valid very near the critical point where the number of states at the split Fermi surfaces are essentially equal.  
Near the critical point, we can expand the integrand in powers of $\bar n/v_Fk_F<<1$, obtaining
\begin{equation}
\bar n= \frac{|F_1^a|}{2}\bar n\left[ C_0(x)-\frac{b}{(v_Fk_F)^2}\bar n^2 C_1(x)\right]+O( (\frac{\bar n}{v_Fk_F})^5),
\label{nx}
\end{equation}
with $C_0(x)\sim 1-(x/2)^{3/2}$,  $C_1(x)=1+x\ln(1+x/5)$ and $F_1^a=f_1^aN(0)$, where $N(0)$ is the density of states at the Fermi surface. 

On the other hand, the integral in eq. (\ref{Delta}) has the  usual ultraviolet divergence of the BCS gap equation. 
A convenient way to deal with this integral  is to sum and subtract $\Theta(\omega_-)$ in the second term and then to subtract the identity 
\begin{equation}
-1+g\int \frac{d^2k}{(2\pi)^2} \frac{1}{2\xi^0}\equiv 0
\label{delta0}
\end{equation}
from the first term of eq. (\ref{Delta}).
$\xi^0=\sqrt{\epsilon_k^2+|\Delta_0|^2}$ and $\Delta_0$ is the uniform
superconducting gap in the absence of Fermi surface splitting.  
With this, eq. (\ref{Delta}) is rewritten as
\begin{equation}
\frac{N(0)}{2}\ln\left|\frac{\Delta_0}{\Delta_q}\right|=
\int \frac{d^2k}{(2\pi)^2} \frac{1}{2\xi}\left\{\Theta(-\omega_+)+\Theta(\omega_-)\right\}\; .
\end{equation}
In this way, the  above integral  is controlled. The coupling constant $g$ and the ultraviolet cut-off
are both contained in the definition of $\Delta_0$ (eq. (\ref{delta0})). 
As before, near the Fermi surface, where interactions are important,
$\Theta(\omega_-)=0$. Performing the remaining integral over $k$, the gap
equation is written  as  
\begin{equation}
\ln\left|\frac{\Delta_0}{\Delta_q}\right|=\Gamma(x)\; ,
\label{gap}
\end{equation}
where $x=\Delta_q/\bar n$, and we have defined the function
\begin{equation}
\Gamma(x)=2\int_{-\bar\theta(x)}^{\bar\theta(x)} \frac{d\theta}{2\pi}
\;\sinh^{-1}\left(\frac{\sqrt{(1+\cos\theta)^2-x^2}}{x^2}\right)\; , 
\label{Gamma}
\end{equation}
with $\bar\theta(x)=\cos^{-1}(x-1)$ for $x\leq 2$ and $\Gamma(x>2)=0$.
Therefore, to look for modulated superconductivity we need to solve eqs. (\ref{nx}) and (\ref{gap}) self-consistently.
Firstly, we note that for $x>2$ the only solution, $\{\bar n=0, \Delta_q=\Delta_0\}$, corresponds to the uniform superconducting phase. 
This imposes a lower limit for modulated superconductivity  since, for obtaining
a non-trivial solution, we must have $\Delta_q<2\bar n<\Delta_0$. This is
consistent with the fact that modulated superconductivity is a strong coupling
effect.
However, for any value of the parameter $\Delta_0> 2\bar n$, the uniform
superconducting mean-field energy is always lower than the modulated one,
$\langle H\rangle_{\Delta_0}<\langle H\rangle_{\Delta_q}$. 
On the other hand, there is a region of the phase diagram, shown above the dashed
line in figure (\ref{pd}), in which $\langle
H\rangle_{\Delta_q}<\langle H\rangle_{\beta}$.  Therefore, modulated
superconductivity appears as a metastable phase for large values of $g$
and $|F_1^a|>2$. The precise location of the onset of metastability depends on
the relation between the band-width in which interactions are relevant (the
energy cut-off), and the band-curvature at the Fermi surface.  
We would like to note that,  the meta-stable character of the modulated SC phase may depend 
on the symmetry of the form factor used to define the effective BCS coupling. 

\section{Discussion and Conclusions}
Summarizing,  we have studied, in the context of Fermi liquids, 
the competition between a Pomeranchuk instability in the spin channel with angular momentum $\ell=1$, 
triggered by the Landau parameter $F_1^a$ and an attractive interaction favoring Cooper-pair formation. 
We built up a zero temperature mean-field phase diagram in terms of the two parameters of the model: 
$F_1^a$ and $g$. We found that, as observed in other channels of 
Pomeranchuk instabilities\cite{metzner-2000,metzner-2007}, 
the superconducting gap reduces the tendency to Fermi surface deformation. 
We have considered  intra-band homogeneous pairing as well as  inter-band
pairing. The net effect of the former is to produce
a $p_y$-wave superconductor state for very small values of the BCS coupling
constant. Then,  we have shown that, for stronger couplings,  a first
order p-wave/s-wave phase transition takes place.
Moreover, we have found a metastable modulated 
superconductor, enhanced for larger values of $|F_1^a|$.

The phase diagram of figure (\ref{pd}) could be modified by  long wave-length
fluctuations in several ways. In the $\beta$-phase, there are three branches of Goldstone modes
transforming with $SO(3)$ group, 
associated with the breaking of relative spin-orbit symmetry. 
The longitudinal mode has a linear dispersion relation, while  the  transverse modes
signal a tendency to produce a non-homogeneous ground state   with parity
symmetry breaking\cite{wu-2007}, meaning that, in the absence of SC interactions, the mean field solution of the 
$\beta$-phase is actually meta-stable. However, an explicit SO coupling ($h$)  can gapped the Goldstone modes, 
stabilizing the phase and  rounding the Pomeranchuk instability.  
On the other hand,  it is known that a modulated superconductor order
parameter   can coexist, at strong coupling,
with non-homogeneous charge configurations\cite{loder-2009}. 
This coexistence has also been  proposed  in the context of
cuprates superconductors\cite{berg-2007}. For this
reason, we believe that transverse long-wave fluctuations of the $\beta$-phase could probably enhance
the modulated SC order.

To the best of our knowledge,  there is still no experimental evidence of the
phases described in this paper. However, we believe that they could be
possibly  observed in systems with  attractive interactions in the antisymmetric $\ell=1$ channel.
Negative values of the Landau parameter  $F_1^a$ have been measure\cite{Legget-1970,Corruccini-1971,Osheroff-1977,Greywall-1983}  in systems like 
$^3He$.  Also, 
 in  ultra cold atomic gases with a p-wave  Feshbach resonance\cite{Zhang-2004},  $F_1^a$ should be negative near the resonance. 
More generally,  strongly correlated systems like cuprates, heavy Fermions and two dimensional electron systems in high magnetic fields
are an interesting arena to investigate the phenomenology presented in this paper.

\section*{Acknowledgments}
D.G.B. is in debt with Eduardo Fradkin for very useful discussions. 
The Brazilian agencies {\em ``Conselho Nacional de Desenvolvimento Cient\'\i
fico e Tecnol\'ogico'', (CNPq)}  and the {\em ``Funda\c c\~ao de Amparo
\`a Pesquisa do Estado do Rio de Janeiro'' , (FAPERJ)}, are acknowledged for
partial  financial support.


\begin{thebibliography}{0}


\bibitem{nozieres-1999} P. Nozieres and D. Pines, {\em The Theory of Quantum Liquids} (Perseus Books, 1999).
\bibitem{pomeranchuk-1958} I. I. Pomeranchuk, {\em Sov. Phys. JETP} {\bf 8}, (1958) 361.
\bibitem{Hiroyuki} H. Yamase and H. Kohno, J. Phys. Soc. Jpn.  {\bf
69}, (2000) 332; J. Phys. Soc. Jpn.  {\bf
69}, (2000) 2151. 
\bibitem{oanesyan-2001} V. Oganesyan, S. A. Kivelson, and E. Fradkin, {\em Phys.
Rev.}
{\bf B64},(2001) 195109 .
\bibitem{barci-2003} D. G. Barci and L. E. Oxman, {\em Phys. Rev.} {\bf B67}, (2003) 205108.
\bibitem{lawler-2006} M. J. Lawler, D. G. Barci, V. Fern\'andez, E. Fradkin, and
L. Oxman, {\em Phys. Rev.} {\bf B73}, (2006) 085101.
\bibitem{Metlitski-2010} M. A. Metlitski and S. Sachdev, {\em  Phys. Rev.} {\bf B82},
(2010) 075127.
\bibitem{barci-2008} D. G. Barci, M. Trobo, V. Fern\'andez, and L. E. Oxman,
{\em  Phys. Rev.} {\bf B78}, (2008) 035114.
\bibitem{zhang-2005} Y. Zhang and S. Das Sarma, {\em Phys. Rev.} {\bf B72},
(2005) 115317.
\bibitem{Hirsch} J. E. Hirsch, Phys. Rev. {\bf B41}, (1990) 6820.  
\bibitem{cabra-2012} P. Rodriguez Ponte, D. C. Cabra, N. Grandi,  {\em Mod.
Phys. Lett.}  {\bf B26}, (2012) 12501.  
\bibitem{wu-2004} C. Wu and S.-C. Zhang, {\em  Phys. Rev. Lett.} {\bf 93},
(2004) 036403.
\bibitem{wu-2007} C. J. Wu, K. Sun, E. Fradkin, and S.-C. Zhang, {\em Phys.
Rev.} {\bf B75}, (2007) 115103.
\bibitem{Sigrist-2004} P. A. Frigeri, D.F. Agterberg, M. Sigrist, New J.\  Phys.{\bf  6}, (2004) 115.
\bibitem{fulde-1964} P. Fulde and R. A. Ferrell,{\em  Phys. Rev.} {\bf 135}, (1964) A550. 
\bibitem{larkin-1964} A. I. Larkin and Y. N. Ovchinnikov, {\em Zh. Eksp. Teor. Fiz.}
{\bf 47},(1964) 1136 , (Sov. Phys. JETP. 20,(1965) 762).
\bibitem{radzihovsky-2008} L. Radzihovsky and A. Vishwanath, {\em Phys. Rev. Lett.} {\bf 103},
 (2009) 010404.
\bibitem{padilha-2009} I. T. Padilha and M. A. Continentino, {\em Journal of Physics:
Condensed Matter} {\bf 21},(2009) 095603.
\bibitem{loder-2009} F. Loder, A. P. Kampf, and T. Kopp, {\em  Phys. Rev.} {\bf B81},
(2010) 020511.
\bibitem{berg-2007}
  E. Berg, E. Fradkin, E.-A. Kim, S. A. Kivelson, V. Oganesyan, J. M. Tranquada, and S. C. Zhang,
   Phys.\  Rev.\  Lett.{\bf  99}, (2007) 127003. 
\bibitem{berg-2009} E. Berg, E. Fradkin, and S. A. Kivelson, {\em  Nature Physics}
{\bf 5}, (2009) 830.
\bibitem{berg-2009b} E. Berg, E. Fradkin, S. A. Kivelson, and J. M. Tranquada,
{\em New J. Phys.} {\bf 11},(2009) 115004.
\bibitem{barci-2011} D. G. Barci and E. Fradkin, {\em Phys. Rev.} {\bf B83}, (2011) 100509.
\bibitem{metzner-2000} C. J. Halboth and W. Metzner, {\em Phys. Rev. Lett.} {\bf 85}, 
(2000) 5162.
\bibitem{metzner-2007} H. Yamase and W. Metzner, {\em Phys. Rev.} {\bf B75},
(2007) 155117.
\bibitem{Zubarev} D. N. Zubarev, Sov. Phys. USPEKHI {\bf 3}, (1960) 320. 
\bibitem{Legget-1970} A. J. Leggett,  J. Phys. {\bf C 3}, (1970) 448. 
\bibitem{Corruccini-1971} L. R. Corruccini, D. D. Osheroff, D. M. Lee, and R. C. Richards,
Phys. Rev. Lett. {\bf 27}, (1971) 650.
\bibitem{Osheroff-1977} D. D. Osheroff, Physica B \& C {\bf 90}, (1977) 20.
\bibitem{Greywall-1983} D. S. Greywall, Phys. Rev. {\bf B 27}, (1883) 2747.
\bibitem{Zhang-2004} J. Zhang, E. G. M. van Kempen, T. Bourdel, L. Khaykovich, J.
Cubizolles, F. Chevy, M. Teichmann, L. Tarruell, S. J. J. M. F. Kokkelmans, and C. Salomon, Phys. Rev. {\bf  A 70}, (2004) 030702.


\end{thebibliography}
\end{document}